\documentclass[pra,twocolumn,showpacs,superscriptaddress,floatfix,nofootinbib]{revtex4-1}
\usepackage{color}
\usepackage{hyperref}  
\usepackage{graphicx}
\usepackage{amsfonts}
\usepackage{footmisc}

\let\a=\alpha \let\b=\beta  \let\g=\gamma  \let\d=\delta 
  \let\h=\eta   \let\th=\theta  \let\l=\lambda
\let\m=\mu    \let\n=\nu         \let\p=\pi    
\let\s=\sigma

\font\tenmib=cmmib10\font\sevenmib=cmmib7\font\fivemib=cmmib5%
\def\BDpr {{\mbox{\boldmath$ \partial$}}}

\def\eqalign#1{\null\,\vcenter{\openup\jot
  \ialign{\strut\hfil$\displaystyle{##}$&$\displaystyle{{}##}$\hfil
      \crcr#1\crcr}}\,}

\def\AA{{\mathcal A}}
\def\EE{{\mathcal E}}\def\DD{{\mathcal D}}

\def\uu{{\V u}}\def\kk{{\V k}}\def\xx{{\V x}}
\def\T#1{{#1_{\kern-3pt\lower7pt\hbox{$\widetilde{}$}}\kern3pt}}

\def\ie{{\it i.e.\ }}
\def\dpr{{\partial}}
\def\defi{{\buildrel def\over=}}

\def\otto{\,{\kern-1.truept\leftarrow\kern-5.truept\to\kern-1.truept}\,}

\newdimen\xshift \newdimen\xwidth \newdimen\yshift \newdimen\ywidth

\def\ins#1#2#3{\vbox to0pt{\kern-#2pt\hbox{\kern#1pt #3}\vss}\nointerlineskip}

\def\eqfig#1#2#3#4#5{
\par\xwidth=#1pt \xshift=\hsize \advance\xshift
by-\xwidth \divide\xshift by 2
\yshift=#2pt \divide\yshift by 2
{\hglue\xshift \vbox to #2pt{\vfil
#3 \includegraphics{#4.eps}
}\hfill\raise\yshift\hbox{#5}}}

\def\V#1{{\bf #1}}
\def\lis#1{{\overline#1}}
\let\wt=\widetilde

\def\tende#1{\,\vtop{\ialign{##\crcr\rightarrowfill\crcr
 \noalign{\kern-1pt\nointerlineskip} \hskip3.pt${\scriptstyle
   #1}$\hskip3.pt\crcr}}\,}
\def\eg{{\it e.g.\ }}

\def\0{\noindent}
\def\*{\vskip2mm}

\def\Eq#1{\label{#1}}
\def\equ#1{(\ref{#1})}

\def\iniz{\setcounter{equation}{0}}
\def\be{\begin{equation}}\def\ee{\end{equation}}
\renewcommand{\theequation}{\arabic{section}.\arabic{equation}}
\newcounter{appendice}
\def\APPENDICE#1{
\setcounter{appendice}{#1}\appendix
\renewcommand{\theequation}{\Alph{appendice}.\arabic{equation}}%
\renewcommand{\thesection}{\Alph{appendice}}%
}

\def\hh{{\bf h}}

\def\alert#1{{\color{ired}#1}}
\definecolor{iblue}{RGB}{65,105,225}
\definecolor{ired}{RGB}{220,20,60}
\definecolor{igreen}{RGB}{50,205,50}
\definecolor{ipurple}{RGB}{75,0,130}
\definecolor{iochre}{RGB}{218,165,32}
\definecolor{iteal}{RGB}{51,204,204} 
\definecolor{imauve}{RGB}{204,51,153}
\let\th=\theta
\begin{document}


\alert{\centerline{\bf
    Ensembles, Turbulence and Fluctuation Theorem}}

\centerline{\bf Giovanni Gallavotti}
\centerline{\small\today}
\centerline{INFN Roma1 \& Accademia dei Lincei}

{\vskip3mm}

\0{\bf Abstract}: The Fluctuation Theorem is considered intrinsically
linked to reversibility and therefore its phenomenological consequence, the
Fluctuation Relation, is sometimes considered not applicable. Nevertheless
here is considered the paradigmatic example of irreversible evolution, the
2D Navier-Stokes incompressible flow, to show how universal properties of
fluctuations in systems evolving irreversibily could be predicted in a
general context. Together with a formulation of the theoretical framework
several open questions are formulated and a few more simulations are
provided to illustrate the results and to stimulate further
checks.\footnote{\email{giovanni.gallavotti@roma1.infn.it}}

\setcounter{section}{0}  
\def\SEC{Introduction.}
\section{\SEC}
\label{sec0}
\iniz
Many macroscopic systems are modeled by equations of motion which are not
reversible due to action of viscous forces, like Navier Stokes fluid
equations. And often the equations can be derived from reversible
microscopic models (\eg see \cite{Ma867-b}). The question that is addressed
in this work is whether stationary states of such systems could also be
described by reversible equations.

In the '80s transport properties of interacting particles systems have been
studied by introducing ``{\it thermostat forces}'',
\cite{No984,EM990,Ho999}: the idea behind the introduction of
``non-Newtonian'' forces was that the important stationary properties of
the system depend on stationarity and not on the way it is achieved. A
feature of resulting equations is that often they are reversible, \ie on
phase-space a map $x\to Ix$, independent of the forces (typically $I$ is
the change of the velocities sign), exists with the property that $I^2x=x$
anticommutes with the time evolution $x\to x(t)=S_tx$ (\ie
$IS_tx=S_{-t}x$).

In the case of equilibrium ergodicity is the basis of the theory of
equilibrium statistical properties and of their independence of initial
conditions; likewise in systems in stationary states (equilibrium or not)
chaoticity of their evolutions takes the role of ergodicity and is the key
to understand the typical initial state independence of the stationary
states properties (with the possibility of a few stationary states, just as
in equilibrium at phase transitions at most a handful of different states
arise depending on the initial conditions), \cite{Ru989b,Ru995}.

This did shew that strongly chaotic evolutions may be
described equivalently by different equations: a striking example is in
\cite{SJ993} in the case of the NS equations and gave rise to several
studies in which strong chaoticity was
present, \cite{Ga997b,Ga002,Ga013b}.

Here a different view will be presented, pursuing the ideas of the latter
references: it is exposed in recent publications and summarized
it in Sec.3-5. The purpose is to investigate whether the proposed analysis
(so far tested only in few cases) can lead to predict properties of
fluctuations that, formally, can be established in reversible systems.

The viewpoint has been that, in the case of systems that are derived
from microscopically reversible mechanical equations (like the NS
equations), the time reversal symmetry being fundamental cannot be broken
and it has to be possible that the same system can be equivalently
described by reversible equations (without, of course, going back to the
atomic scale description of the motion).

If so it is natural to ask if certain properties of reversible chaotic
evolutions can be manifest in irreversible ones, chaotic or not, as the
chaos of the microscopic motions at the base of the macroscopic ones will
be always active (even when the macroscopic motion is not chaotic, \eg in
NS at very low Reynolds number).

The general ``equivalence conjecture'' is presented in Sec. 3-5 is applied,
assuming the validity of the ``Chaotic Hypothesis'' (CH),
\cite{GC995,Ga019a}, to predict fluctuations of the dissipation (defined as
phase space volume contraction rate) in a simple incompressible NS equation
in 2D with periodic b.c. and constant viscosity.

The (new) result is presented in Sec. 7, where a positive test is analyzed.
In Sec.8 it is argued that the test should be regarded only as a first step
because simulations deal with cases in which the motion seems to involve an
attractor that fills the available phase space. The really interesting case
occurs at high Reynolds number, \ie at small viscosity, where the attractor
is really a tiny surface in phase space: from the results of Sec. 7 it
appears that much more stringent test are suggested for future work and are
possible with the available computer resources.

The equivalence conjecture is intriguing also because, not only it suggests
a close analogy between the theory of equilibrium ensembles and the theory
of the NS equation stationary states, but it applies as well to 3D while it
does not conflict with the possible existence of singularities in 3D-NS: it
deals with properties of the NS evolution {\it regularized} by a UV cut off
which become {\it independent} on the regularization scale if large enough.

A few new tests of the equivalence are also included as supplementary
material, with attention to properties of the Lyapunov spectrum not covered
by the equivalence conjecture to which it nevertheless appears,
surprisingly, closely related. Of particular interest will be the study of
the 3D case: some work on it is being done and a few papers (quoted below)
are appearing.

\def\SEC{Navier-Stokes flow in 2D. Notations.}
\section{\SEC}
\label{sec1}
\iniz
Here we consider the simple case of an incompressible fluid in a periodic
container in dimension $2$ (2D), review \cite{Ga019d,Ga997b}. Velocity
$\uu(\xx)$ can be expressed via a Fourier's series; if the container side
is $2\p$ then:
\be\uu(\xx)=\sum_{\V0\ne\kk=(k_1,k_2)\in Z^2} u_\kk e(\kk)
  e^{-i\kk\cdot\xx}, \qquad e(\kk)\cdot \kk=0
\Eq{e1.1}\ee
with $u_\kk=\lis u_{-\kk}$ scalars, $\kk^\perp=(k_2,-k_1)$,
$e(\kk)=\frac{i\kk^\perp}{|\kk|}$

The NS equation for the components $u_{\kk}$ is then:
\be\eqalign{&\dot{u}_\kk=\EE(u)_\kk-\n \kk^2u_\kk+f_\kk, 
\cr
&\EE(u)_\kk=-\kern-3mm\sum_{\kk_1+\kk_2=\kk}
\frac{(\kk^2_2-\kk^2_1)\,(\kk_1^\perp\cdot\kk_2)}{2 |\kk_1||\kk_2||\kk|}
u_{\kk_1} u_{\kk_2},\cr}\Eq{e1.2}\ee

Forcing will be supposed to ``act on large
scale'': $f_\kk\equiv\V0$ for $|\kk|>K$ for some $K$. It is convenient to
imagine that $f$ is fixed {\it once and for all} and $\sum_\kk|
f_\kk|^2=1$: below the case $f_\kk=\lis f_{-\kk}\ne0$ only for
$\kk=\pm\kk_0, \kk_0=(2,-1)$ and random phase will be considered.

Hence the {\it only dimensionless parameter} in the NS equation to which,
for brevity, we refer as the ``Reynolds number'', is $R\equiv\n^{-1}$. The
NS equations will be considered with ultraviolet regularization $N$, \ie
Eq.\equ{e1.2} in which all $\ne\V0$ waves $\kk,\kk_1,\kk_2$ have components
of modulus $\le N$. Of course we are interested in properties which {\it do not
depend on $N$} at least for large $N$.

Notable cancellations are expressed by
the identities:

\be {\sum}_\kk \lis {u_\kk}\EE(u)_\kk=0, \quad
{\sum}_\kk\kk^2 \lis {u_\kk}\EE(u)_\kk=0\Eq{e1.3}\ee 
which, in the 2D incompressible Euler flow with no stirring (\ie $\n=0,f=0$),
imply conservation of energy $E$ and enstrophy $\DD$ (\ie
  $E=\sum_\kk |\uu_k|^2, \DD=\sum_\kk |\kk|^2|\uu_\kk|^2$).  The identities
Eq.\equ{e1.3} remain valid even in presence of the UV cut-off, \ie if all
$\ne\V0$ components of $\kk,\kk_1,\kk_2$ in Eq.\equ{e1.2},\equ{e1.3} are
restricted to be $\le N$.

\def\SEC{Ensembles and nonequilibrium fluids}
\section{\SEC}
\label{sec2}
\iniz

In Statistical Mechanics (SM) equilibrium states of a system can be {\it
equivalently} described by a probability distribution in different
ensembles (canonical, microcanonical and others). In the
review \cite{Ga019d} an analogous paradigm (evolved from the earlier work
\cite{Ga997b,Ga019d,Ga019a}) has been proposed to hold for
stationary states in fluid mechanics (actually in more general stationary
nonequilibria).

The idea (already presented in the earlier publications, mostly in the form
of a proposal for a project) is that in the NS equation the same mechanism
that is well known in SM could (should) operate: namely there are different
probability distributions which assign the same averages to a large class
of observables, \ie the ``{\it local} ones depending only on the particles
that happen to be in a fixed region $L$ of space, as long as the region is
small compared to the total container volume $V$ (ideally infinite).

The proposal is to identify, in NS, the ``local observables'' with the
functions of the velocity field which depend only on the components $u_\kk$
with $|\kk| < K$ with $K$ small compared to the UV cut off $N$ that regularizes
the equation (necessary in the 3D case). So $N$ plays the role of the total
volume $L$ and $K$ the role of the (arbitrarily fixed) finite volume (physically
$K\ll L$).

As in SM completely different distributions describe the same system
provided their parameters are suitably fixed. For instance, fixing density
$\rho=1$, you can use the microcanonical distrbution with energy $E = e V$ or
the isokinetic distribution with kinetic energy $T = \frac2{3\beta}V$ and the
two distributions assign {\it exactly} the same averages to the local
observables, provided the constant microcanonical value of the energy equals
the average value of the energy in the isokinetic distribution; or,
in the microcanonical and canonical equivalence, the total microcanonical
energy equals the average canonical energy.

In the following we consider two evolution equations for the fluid.
\*
1) Denote $t\to S^{irr,N}_t\uu=\uu(t)$ a solution to the NS equations with
UV cut-off $N$. {\it Time reversal} $I\uu=-\uu$ is not a symmetry, \ie $I
S^{irr,N}_t\ne S^{irr,N}_{-t}I$, because of viscosity $\n>0$.

2) Consider also Eq.\equ{e1.2}, with the same UV regularization:
\be \dot u_\kk= \EE(u)_\kk-\a(u)\kk^2 u_\kk+f_\kk\Eq{e2.1}\ee
but with viscosity $\n$ replaced by a multiplier $\a(u)$ designed so that
the enstrophy $\DD(u)=\sum_\kk |\kk|^2|\uu_\kk|^2$ {\it is conserved}.
denote $t\to S^{rev,N}_t\uu=\uu(t)$ the evolution for Eq.\equ{e2.1}.
\*

In 2D the second Eq.\equ{e1.3} yields:
\be\a(\uu)=
\frac{\sum_\kk \kk^2 \lis f_\kk \uu_\kk}{\sum_\kk |\kk|^4 |\uu_\kk|^2}
\Eq{e2.2}\ee
which also immediately implies that flows $t\to \uu(t)=S^{rev,N}_t\uu$ of
Eq.\equ{e2.1} are {\it reversible} \ie $I
S^{rev,N}_t=S^{rev,N}_{-t}I$.\footnote{\small In 3D the second of
  Eq.\equ{e1.3} does not hold: see appendix B for the 3D version of
  Eq.\equ{e2.2}.}

At fixed forcing $f$, for each choice of the control parameters, \ie
$R=\n^{-1}$ for the irreversible $S^{irr,N}_t$ and $D$ the enstrophy
constant for the reversible $S^{rev,N}_t$, a unique stationary distribution
$\m^{irr,N}_R$ or $\m^{rev,N}_D$ is determined, if $\n$ is mall enough;
which yields the statistical properties of the stationary state reached
from (volume)-almost all initial $\uu$ in the phase space $M^N$. At small $R$
there might be different stationary states that can be reached with
positive probability depending on the initial data $\uu$. All such
distributions will be collected in ``ensembles'': $\EE^{irr}$ or $\EE^{rev}$
rspectively.

The goal is to see whether a $1-1$ correspondence between the distibutions,
in the irreversible ensemble $\EE^{irr}$ and in the reversible one
$\EE^{rev}$, can be established so that in the limit $N\to\infty$, \ie
removing the UV cut off, corresponding distributions assign the same average
value to the local observables.

Existence of a correspondence with the latter property will be called {\it
  Equivalence Hypothesis}.

\def\SEC{Equivalence}
\section{\SEC}
\label{sec3}
\iniz

The well known difficulty with achieving control of the enstrophy is to be
expected to correspond to an evolution of $\a(S^{rev,N}_t\uu)$ with extreme
fluctuations at least at large Reynolds number
$R=\frac1\n$.\footnote{\small In 2D ({\it only}) enstrophy can be
  controlled but it can grow up to $\n^{-2}=R^2$, \cite[Eq3.2.24]{Ga002}.}

This in turn might produce a {\it homogeneization} phenomenon which could
imply that $\a$ can be replaced, for practical purposes, by a constant:
leading to statistical properties similar to those of the irreversible
evolution $S^{irr,N}_t$, at least on {\it local observables}, \ie
observables $O(\uu)$ depending on $\uu$ via its Fourier components $u_\kk$
with $|\kk|< K$ with $K$ fixed (arbitrarily) but $\ll$ than the UV cut off
$N$.

Possibility of equivalent descriptions of stationary states of
turbulent fluids arose in the key work \cite{SJ993}: where the NS
equation has been shown to be describable, in simulations, by the
stationary state of a different fluid equation obtained by imposing on the
Euler equation the constraint that the energy content of ``each shell'' in
$\kk$-space is set to the value predicted by the $5/3$-law.

In the latter reference, at difference with Eq.\equ{e2.1}, the constraint
was imposed via as many multipliers as inertial shells: yet it led to
reversible equations of motion which, remarkably, were shown to attribute
to several large scale observables averages (\ie local observables in the
above sense) sharing the statistical properties obtained from the
corresponding irreversible NS.

The following equivalent ensembles description is proposed,
see \cite{Ga019d} for a review, for the stationary states of the
incompressible fluid.

Let $\EE^{irr,N}$ be the family of stationary distributions that can be
reached by evolving, via the usual NS Eq.\equ{e1.2}, initial velocity
fields $\uu$ selected with probability $1$ with respect to {\it (any)}
distribution with density $\rho(\uu)d\uu$ on the phase space $M^N$
defined\footnote{If the UV cut-off is intended as contraining all
  components of $\kk\ne\V0$ to be $|k_i|\le N$, then the real dimension of the
  space $M^N$ is $4N(N+1)$ if $d=2$, as for each $\kk\ne\V0$ there is one
  complex coordinate, and $\uu_\kk=\lis{{\uu_{-\kk}}}$.} by the Fourier's
coefficients $\uu_\kk$ of $\uu$. The conceptual importance and the role of
the selection criterion, adopted here, has been stressed and used by
Ruelle, see the reviews \cite{Ru989b,Ru995} and \cite{Ga019d}.

Existence of the stationary states will be, here, a consequence
of a general assumption, {\it Chaotic Hypothesis}, on systems which are
``chaotic'' (\ie have some positive Lyapunov exponents), supposed to hold
throughout.\footnote{The formulation goes back to \cite{GC995},
for a review see \cite{Ga013b,Ga019d}:
{\bf Chaotic Hypothesis (CH):\it Evolution of a chaotic system is attracted
by a smooth surface in phase space and, on it, it is a smooth Anosov
system.}
``Anosov systems'' are too often still misunderstood (or criticized as
constructs by ``some mathematicians'', \cite[p.219]{Ho999}) even though the
works \cite{Si968b,BR975} have been popularized in many later publications
by the same authors. Still they are the simplest general examples of
chaotic motions and should be regarded to play, in chaotic dynamics, the role
played in non-chaotic dynamics by the harmonic oscillators.  CH implies,
\cite{Ru010}, existence of a unique stationary state associated with each
attractor: it is a ``genericity'' hypothesis and here it is supposed to
hold for the evolutions considered. It is an interpretation of the (weaker)
hypothesis that motion near the attractors is a Axiom A system,
\cite{Ru989b,Ru995}.}

At small viscosity $\n=\frac1R$, \ie large Reynolds number $R$, it is
expected that there is a unique stationary state (\ie a probability
distribution for the local observables) $\m_R^{irr,N}(d\uu)\in\EE^{irr,N}$:
discussing the (well known, \eg \cite{FGN988}) possible non uniqueness will
also be considered later below.

Likewise let $\EE^{rev,N}$ be the family of stationary distributions that
can be built in the same way via Eq.\equ{e2.1}: the distributions can be
parameterized by the value of the enstrophy $\DD$, which is a constant
$D=\DD(\uu)$, fixed by the initial datum enstrophy. And for large $D$ it is
expected that there will be a unique stationary state
$\wt\m^{rev,N}_D(d\uu)\in\EE$.

In \cite{Ga019a} it is proposed, ``{\it Equivalence Hypothesis}'' (EH for
brevity), that in a turbulent regime (\ie at small $\n$ or large $D$) the
above $\m^{irr,N}_\n$ for the irreversible flow and $\wt\m^{rev,N}_D$ for
the reversible will be {\it equivalent as $N\to\infty$ if}
\be \m^{irr,N}_\n(\DD)=D_N\Eq{e3.1}\ee
\ie {\it if the enstrophy $D_N$ in $\wt\m^{rev,N}_{D_N}$
is the irreversible evolution
average of the enstrophy $\DD(S^{irr,N}_t\uu)$}. The $D_N$ in the r.h.s
will in general depend on $N$: remark, however, the {\it a priori} bound
$D<R^2$ for all $N$ valid in 2D (due to Eq.\equ{e1.3}).

The precise meaning is that, fixed $\n$, for {\it any local observable} $O(\uu)$
(\ie of large scale, as defined in paragraph after Eq.\equ{e2.2}) it will
be, under condition Eq.\equ{e3.1}:
\be\lim_{N\to\infty} \m^{irr,N}_\n(O)=\lim_{N\to\infty} \wt\m^{rev,N}_{D_N}(O)
\Eq{e3.2}\ee
This will be briefly denoted $\m^{irr,N}_\n\sim\wt\m^{rev,N}_{D}$.

The analogy with the {\it equivalence in SM} between 
canonical and microcanonical ensembles is stressed in \cite{Ga019a}: with
$\n,D,N\to\infty$ playing the role of $\b,E,V\to\infty$ (inverse
temperature, energy and 'thermofynamic' limit ).

\def\SEC{Work per unit time}
\section{\SEC}
\label{sec4}
\iniz

It is worth spending a few words on the energy balance in stationary
states: it provides important insights on the equivalence hypothesis (EH).

The work of the stirring force per unit time $W=\sum_\kk f_\kk \lis u_\kk$
is a local observable, by the assumption that $f_\kk=0$ unless $|\kk|<K$
for some fixed $K$ (see paragraph following Eq.\equ{e1.2}).

Hence the implication of EH, \ie that Eq.\equ{e3.1} implies Eq.\equ{e3.2},
yields:
\be \m_R^N(W)=\wt\m^{rev,N}_{D_N}(W)\Eq{e4.1}\ee
This is obtained  by just multiplying by $\lis \uu_{\kk}$ both sides of the
equations Eq.\equ{e1.2} (irreversible NS equation) and Eq.\equ{e2.1}
(reversible NS) and summing over $\kk$: with the result
\be \frac{d}{dt} \frac12\sum_\kk |\uu_\kk|^2= -\g \DD(\uu)+ W(\uu)
\Eq{e4.2}\ee
where $\g=\n$ for $NS^{irr,N}$ or $\g=\a(\uu)$, for $NS^{rev,N}$ (the
inertial terms cancel\footnote{\small This remains true in 3D NS.}
exactly, Eq.\equ{e1.3}). Averaging, under the
condition Eq.\equ{e3.1}, over time gives in the two cases:
\be \eqalign{&\m^{irr,N}_\n(W)-\n
\m_\n^{irr,N}(\DD)=0,\cr
&\m^{rev,N}_{D_N}(W)-D_N\m^{rev,N}_{D_N}(\a)=0\cr}\Eq{e4.3}\ee
Thus the physically appealing Eq.\equ{e4.1}, consequence of the
equivalence hypothesis, provides an important {\it test} of it via the
energy balance in 
Eq.\equ{e4.3}. Namely it implies that the multiplier $\a$ in Eq.\equ{e2.2} has
an average $=\n$:
\be \n=\lim_{N\to\infty}\m_{D_N}^{rev,N}(\a), \ \ie\
\lim_{N\to\infty} R \m_{D_N}^{rev,N}(\a)=1\Eq{e4.3}\ee
thus allowing the interpretation of the equivalence in terms of a
``homogeneization property'', as proposed above. It supports the suggestion
that equivalence relies on chaotic evolution of the multiplier $\a$ and
leads to a first nontrivial test of equivalence: \ie Eq.\equ{e4.2} follows
from the equivalence condition Eq.\equ{e3.1}.

This also shows that the equivalence hypothesis could be also formulated
replacing Eq.\equ{e3.1} with $\n_N=\m^{rev,N}_D(\a)$ (this time $\n$ will
depend on $N$ as $D$ did in Eq.\equ{e3.1}) and, in this case, the relation
$\lim _{N\to\infty}\m^{irr,N}_{\n_N}(\DD)=D$ would be a nontrivial test.

The above analysis establishes a $1-1$ correspondence between the elements
of the distributions in $\EE^{irr}$ and $\EE^{rev}$ of stationary states
of the two equations in the region of parameters $\n,D$ in which the
equations admit a unique stationary distribution (with probability $1$ with
respect to the choice of initial data with a distribution with density on
phase space, called SRB-distributions).

However it is known that often, even with fixed and constant forcing, as it
is the case here, the evolution may be attracted by different
attracting sets, each with a probability $>0$, particularly if $\n$ is
large (at fixed $N$ and small Reynolds number), \cite{FGN988}.

The hypothesis should then be extended. A natural extension is that the set
of extremal (\ie ergodic) stationary states with given $\n$ or $D$ are in
$1-1$ correspondence and each pair $\m^{irr}_{\n,\h}, \wt\m^{rev}_{\h,D}$,
labeled by an extra index $\h$, is reached as a limit as $N\to\infty$ of
$\m^{irr,N}_{\h,\n}, \wt\m^{rev,N}_{\h,D}$ (which might depend on the
initial data or even on alternative ways of realizing the UV cut-off).
See \cite{Ga019a}: a situation analogous to that arising in the theory of
phase transitions in Statistical Mechanics, \cite{Ru969}, see the related
analysis in \cite{SDNKT018}.

\def\SEC{Equivalence tests}
\section{\SEC}
\label{sec5}
\iniz

Tests of equivalence can be found in several publications; to
mention the most recent: \cite{Ga002,Ga013b,Ga018,Ga019a,Ga019c,Ga019d}.

As an example a test of the key relation Eq.\equ{e4.3} is reported in Fig.1.
Fixing the viscosity a
$\n=\frac1R=1/2048$ and $N=31$, \ie $3968$ modes:
\*\*\*\*

\eqfig{230}{140}{}{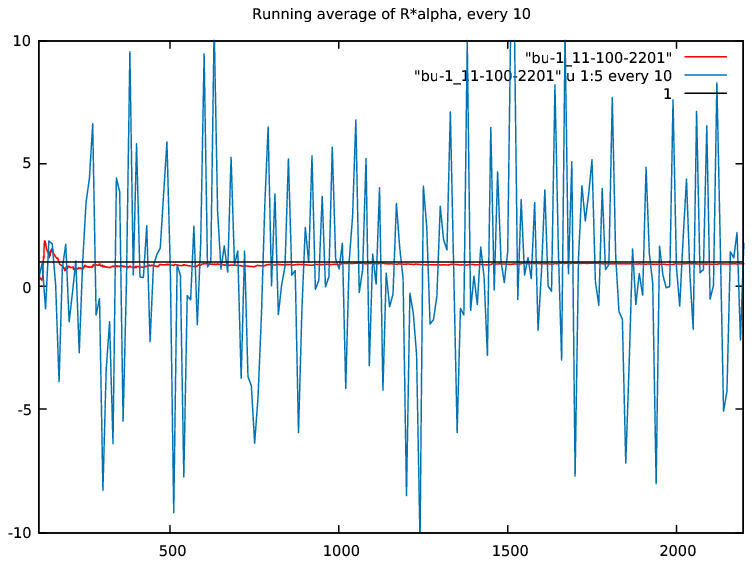}{}

\vskip-3mm \0{\small Fig.1: The axis is time in units $2/h$ with
  $h=2^{-14}$ as integration time step, with Runge-Kutta-4 integrator. The
  (blue o.l.) fluctuating line yields the time evolution of the multiplier
  $R\a(t)$ (Eq.\equ{e4.3}) in the {\it reversible evolution} ($NS_{rev}$);
  the (red o.l.) line yields at each time $t$ the time average between the
  initial time up to $t$, which should be a line asymptotic to $1$, which
  is reached within $10\%$ amid fluctusations $150$ times as large in a
  relatively short (due to computer time constraints) run.  And the
  horizontal line, a visual aid, is the line at height $1$. The total run
  is over $t\in[0,2200]$ with the time unit which is $2^{15}$ integration
  steps each of size $2^{-14}$; and the initial data are random while the
  forcing has only one complex mode, namely $\kk=\pm(2,-1)$. Here
  $R=2048,N=31$, $3968$ modes (the key numbers are 1=theoretical prediction
  of the rev.evol. average of $R\a(t)$, $11=\log_2
  R$, $100$=initial time, $2200$=final time).}  \*
  
Fig.1 has been obtained via a semispectral code: in a non spectral method
this should be comparable to a $63^2$ discretization.  Remarkably the same
simulation, see \cite[fig.2]{Ga019c}, can be done measuring the multiplier
$\a(\uu)$ in the irreversible $NS_{irr}$ evolution, regarding it as an
observable defined by Eq.\equ{e2.2}: this is not a local observable, still
the result is very close to the one in Fig.1. In this case, although $\a$,
regarded as an observable for the irreversible $NS_{irr}$ flows, is non
local still, in corresponding distributions, its running average has the
same average in $NS_{rev}$ and $NS_{irr}$. This hints at the possible
existence of families of non local observables which fall into the
equivalence: a point on which to return below.

The same simulation for $NS_{rev}$ can be performed at much larger
friction, \eg smaller $R\simeq28$, and just $48$ modes. This time the
phenomenology is somewhat different and the variable $\a$ undergoes much
smaller fluctuations becoming only rarely negative.
Equivalence is however respected: increasing viscosity the multiplier $\a$,
while strongly fluctuating, will much less fluctuate relatively to its
average. Eventually at very large viscosity the flow, in the stationary
states, becomes laminar or periodic and fluctuations of $\a$ no longer
extend to negative values.

Finally it should be stressed that the 2D nature of the equations is {\it not
essential} and all the general ideas carry {\it unchanged} to 3D: in particular
the question of existence and uniqueness of the NS equation in 3D does not
arise: the ``only'' difference is that attention should be really paid to
the $N$ dependence of $D$ in Eq.\equ{e3.1}, no longer constrained by the
mentioned {\it a priori} upper bound. Studies of the 3D reversible NS
and its relation with the 3D irreversible are beinning to appear: see
\cite{SDNKT018,AB020} for NS and \cite{BCDGL018} for the shell model.

\def\SEC{Fluctuation Theorem}
\section{\SEC}
\label{sec6}

After the above introduction, summarizing earlier work, consider next the
main new question studied in this note.  Assuming equivalence it is natural
to ask whether reversibility of the $NS_{rev}$ evolution gives new insights
in the corresponding $NS_{irr}$ irreversible flows.

Consider the Fluctuation Theorem (FT): for reversible Anosov systems it
deals with the phase space contraction (physically interpreted as ``entropy
production rate'', \cite{Ga019a}) whose fluctuations exhibit universal
properties.

In the $NS_{rev}$ evolution the non constant multiplier $\a$ leads to a
``phase space contraction'' (\ie the ``divergence'', formally $\sum_\kk
\frac{\dpr \dot u_\kk}{\dpr u_\kk}$) which, after a brief calculation, is:
\be\s(u)=\a(u)\Big(2K_2-2\frac{E_6(u)}{E_4(u)}\Big)+\frac{F(u)}{E_4(u)}
\Eq{e6.1}\ee
with $\a$ in Eq.\equ{e6.1} and $K_2,E_4(u),E_6(u),F(u)$ are:
\be \eqalign{
  2K_2=&\sum_\kk \kk^2, \ E_4(u)=\sum_\kk(\kk^2)^2 |u_\kk|^2,\ 
  \cr 
E_6(u)=&\sum_\kk(\kk^2)^3 |u_\kk|^2,\
F(u)=\frac{\sum_\kk (\kk^2)^2\lis f_\kk
u_\kk}{E_4(u)}
\cr}\Eq{e6.2}\ee
where the sums run over the $\kk$ with $|k_i|\le N,\, i=1,2$.

In {\it time reversible Anosov systems} the fluctuations of the divergence
satisfy a general symmetry relation. Namely if $S_t$ denotes the evolution
and $\s_+$ the infinite time average of $\s(S_t u)$ and
\be p(u)=\frac1\tau\int_0^\tau
\frac{\s(S_\th u)}{\s_+}d\th \Eq{e6.3}\ee
then $p$ has a probability distribution in the stationary state
such that $p\in dp$ has density $P(p)=e^{s(p)\tau +O(1)}$,
asymptotically as $\tau\to\infty$, 
with the {\it universal} symmetry, \cite{GC995}:
\be s(-p)=s(p)-p \tau \s_+\Eq{e6.4}\ee
called the ``Fluctuation Theorem'' (FT).

In applications it would be important to know that Eq.\equ{e6.4} holds:
however in any laboratory experience the relation {\it cannot} be
considered mathematically satisfied because it is essentially impossible to
check both the CH {\it and} the reversibility.

Several attempts can be found to study empirically the relation
Eq.\equ{e6.4} which, when it cannot be {\it a priori} proved, is called
``Fluctuation Relation'' (FR).

Before posing the main question: {\it is it meaningful to ask whether the FR
holds in irreversible evolutions ?} it is necessary studying, first:

1) the probability distribution $P$ of $p$, defined by Eq.\equ{e6.3} both
in the reversible and irreversible flows: although $p$ is not a local
observable,  nevertheless it might be among the non local observables with
equal or close corresponding distributions, like the $R\,\a$ illustrated in
Fig.1, \cite{Ga019a,Ga019d}.

2) the {\it local Lyapunov spectrum}: defined by considering the Jacobian matrix
of the evolution, formally the matrix $J_{\kk,\hh}=\frac{\dpr \bf \dot {\rm
    u}_\kk} {\dpr u_\hh}$, and then computing the eigenvalues of its
symmetric part, in decreasing order, and averaging each one over the
flow, see \cite[p.291]{Ru982}.

Whether the spectra of the reversible and irreversible evolutions are
close is related to the key question: because in reversible
Anosov systems the number of exponents $\ge0$ equals that of negative
exponents. Hence their equality {\it indirectly tests CH}.

Preliminarily it should be asked whether the FR is even to be expected at
least for the stationary flows obeying reversible $NS_{rev}$. The CH, which
is assumed, will imply that the evolution is a Anosov flow on the
attracting surface.  However, to apply the theorem, evolution should also
be time reversible: and if the attracting set is not the full phase space
the FT cannot be applied, at least not without further
work.\footnote{\small If the attracting surface $\AA$, see CH, is not the
  full phase space $M^N$ then the time reversal image $I\AA$ is likely to
  be disjoint from $\AA$ and the motion restricted to $\AA$ is not
  symmetric under the natural time reversal $I$.\label{AA}}
 
Hence a simple check will be to count the numbers of positive and negative
exponents: if the negative ones are more than the non negative the
evolution {\it on the attracting manifold} cannot be reversible in spite of
the time reversibility of $NS_{rev}$ on the full phase space.

The test turns out to be possible, in a reasonable computer time, in the
simple case of the NS equation with very few modes, $48$ modes and $R=2048$
(a modest $7\times7$ grid): the above defined local Lyapunov spectrum can
be equivalently defined as the Lyapunov exponents of the trivial linear
flow $S_tv = e^{J^s(\uu)t} v$: therefore it can be computed either using a
packaged routine for the computation of eigenvalues or using (as done here)
the algorithm in \cite{BGGS980b}.\footnote{\small Fast in this case, if the
  time series $\uu(t)$ is available (which is provided by the simulations
  needed to draw graphs like Fig.1) because $\uu$, hence, $J(\uu)$ remain
  fixed: they are here computed by iterating a large number $k$ of times
  (of the order of $h^{-1}$) the matrix $(1+h J^s(\uu))$ and applying the
  quoted method. To obtain $k$-independent results the time series should
  be taken at time intervals large enough. Within the graph one should
  recognize one exponent $0$ in the $NS^{irr}$ and two in $NS^{rev}$: but
  the relative errors are large precisely near $\l_k=0$ and blur this
  property (to visualize the error sizes in the problem see Fig.8 of the
  Appendix A, and Fig.6,7,8, which illustrate the almost
  identity of the average local spectra amid impressive fluctuations
  particularly in the reversible evolutions).}

\eqfig{180}{115}{}{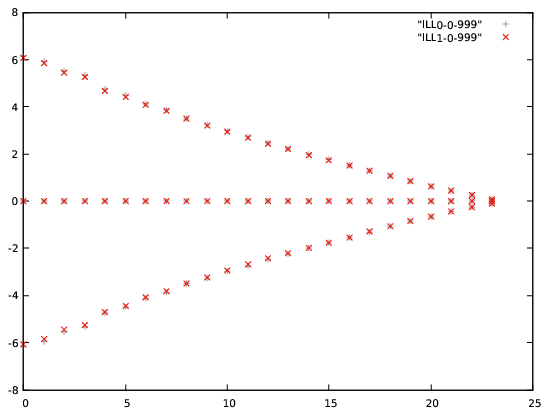}{}
%
%
\vskip-2mm \0{\small Fig.2: Local Lyapunov spectra for {\it both}
  $NS_{irr}$ {\it and} $NS_{rev}$ flows with $d=48$ modes, $R=2048$. Rapid
  computation with only $1000$ samples taken every $4/h$ time steps of time
  $h=2^{-13}$ and averaged: the upper and lower values give the $d/2$
  exponents $\l_k$ and respectively $\l_{d-1-k}$, while the middle values
  are $\frac12(\l_k+\l_{d-1-k})$ not constant but close to $\simeq
  -.01$. This figure shows positive exponents to be equally numerous as the
  negative ones and the features {\bf a),b)} listed below. } \*

The quick check in Fig.2 (see also \cite{Ga019a,Ga019c}) reports
$\l_k,k=0\ldots d/2-1$: the first half of the $d=4N(N+1)$ exponents in
decreasing order and the second half $\l_{d-1-k},k=0\ldots d/2-1$ as
function of $k$ (upper and lower curves), as well as
$\frac12(\l_k+\l_{d-1-k})$ (intermediate line). 

It yields other somewhat
surprising results besides showing the equality of the numbers of positive
and negative exponents which, as mentioned above, we take as evidence that the
attracting set fills densely phase space so that the time reversal symmetry
remains a symmetry on the attracting set.  Figure draws
{\it in the same panel}, spectra from both $NS_{rev}$ and $NS_{irr}$ flows
under equivalence conditions; they apparently ovelap and show: \\
{\bf a)} ``coincidence'' of the spectra of the $NS_{rev}$ and $NS_{irr}$
evolutions: quite surprising and justifying an attempt to formulate and
check the Fluctuation Relation in the {\it irreversible} flows.  \\
{\bf b)} apparent ``pairing'': the exponents appear ``paired'', \ie
$\frac12(\l_k+\l_{d-1-k})$ is $k$-independent. Further results on pairing
in the Appendix.  Therefore the flow, being reversible and having equal
number of pairs of opposite sign, can be consistently assumed to be a
Anosov flow and
\\
{\bf c)} The local Lyapunov spectrum is related to the actual Lyapunov spectrum
via interesting inequalities, \cite{Ru982,Li984}: which could be used to
test accuracy of simulations.\footnote{\small The inequalities do not
  estimate the number of positive exponents proportionally to the
  enstrophy, not even in dimension $2$; a question is whether such a bound
  could hold in dimension $2$ by allowing a proportionality constant $(\log
  N)^c$ for some $c$.}
\*

The compatibility of the latter result with reversibility suggested to test
the FT. The graph for $(s(p)-s(-p))/\s_+\tau$ in Eq.\equ{e6.4} is studied for
both reversible and irreversible flows. It exhibits the main result of this
work: \*\*

\eqfig{180}{130}{}{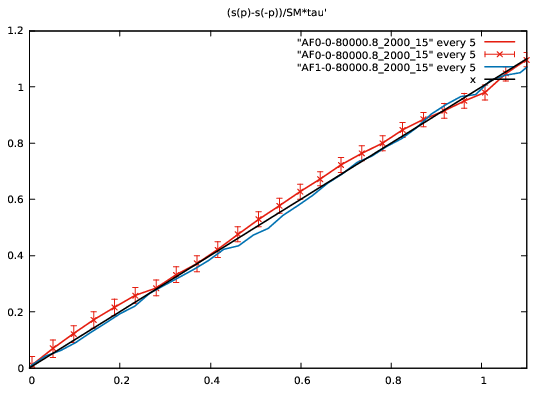}
.\hglue.5cm\raise 2cm \hbox{\kern3.5cm\tiny A}
\vskip-2mm \0{\small Fig.3: Test the fluctuation relation in the flow
  $NS_{irr}$ (red o.l.) and $NS_{rev}$ (blue o.l.) flows with $48$ modes,
  $R=2048$. The $\tau$ is chosen $8$, the slope of the graph increases with
  $\tau$ reaching $1$ at $\tau=8$. The graph is built with $8\cdot 10^4$ data,
  divided into $2\cdot10^3$ bins, obtained sampling the flow every $4/h$
  time steps of size $h=2^{-13}$. The keys AF0 and AF1 deal with $NS_{irr}$
  (red o.l.) and, respectively, $NS_{rev}$ (blue o.l.)  and the error bars
  (red o.l.) deal with $NS_{irr}$; the line $f(x)=x$ is a visual aid.}
\*

The histogram of the PDF corresponding to Fig.3 is very close to a Gaussian
centered at $1$ and width yielding the slope of Fig.3: 
\*
\eqfig{180}{115}{}{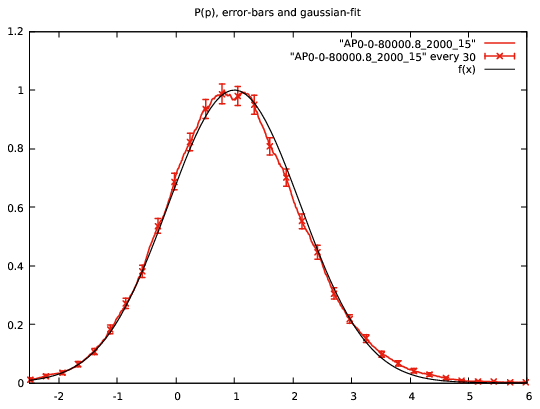}
.\hglue.5cm\raise 2cm \hbox{\kern3.5cm\tiny A}
\vskip-2mm \0{\small Fig.4: A histogram (with max normalized to $1$), of
  the PDF {\it for the irreversible flow} of the variable $p$ (red o.l.),
  with $\tau=8$ generating the Fig.3 out of the $8\cdot 10^4$ measurements
  of $\s(\uu)$ in the $NS_{rev},NS_{irr}$ equations. The $p$-axis is
  divided in $2000$ bins and for each $p$ the average of the number (and
  corresponding error bars) of points in $[p-\d,p+\d]$ is plotted (red
  o.l.)  with $\d=15/2000$ (corresponding to a small interval of $p$
  compared to the width $2\sqrt{\s_+\tau}$) and the interpolating Gaussian
  (blue o.l.). The error bars for the reversible flow (not drawn) have the
  same sizes.}  \*

Fig.3 also shows that the proposed equivalence {\it extends} also to the
phase space contraction (``entropy production rate'', \cite{Ga013b,Ga019a})
as a quantity defined for the reversible evolution but regarded as an
observable for the irreversible $NS$. The interest of the result in Fig.3
is to provide an {\it a priori} predicted fluctuation relation in a {\it
  irreversible} evolution.\footnote{\small In summary the prediction is
  based on CH, on the equality of numbers of negative and non negative
  exponents and on the extension of the equivalence hypothesis to the
  entropy production rate.}

\iniz
\def\SEC{Problems on strong dissipation}
\section{\SEC}
\label{sec7}

The results on the fluctuation relation (FR) are very special because
the UV regularization is so small that the number of (local) Lyapunov
exponents can be easily computed and checked to be the same for negative 
and nonnegative ones. This makes possible to suppose that CH holds and that
the attracting surface is the entire phase space, so that time reversal is a
symmetry for the evolution on the attractor: which implies that the FR
follows from the FT and leads to the above test.

{\it More interesting} would be the case of higher regularization: already
at $224$ modes the number of negative exponents {\it exceeds} that of the
positive ones. The first remark is that, nevertheless, the (approximate)
``pairing'' between exponents already quite clear in Fig.2 remains a
characteristic feature, as the cut-off $N$ increases, see Fig.5 below.

Three objections can be raised, before even beginning to attempt possible
application of the FT to NS evolutions with strong dissipation and several
momentum scales.

1) excess of negative Lyapunov exponents which indicates (if CH holds) that
the flow evolves towards an attractor of dimension smaller than the full
dimension of phase space: this breaks time reversal symmetry which ceases
to be a symmetry of the evolution on the attractors (although it {\it
  remains} a global symmetry for the $NS_{rev}$ flows).

2) if the attracting set dimension is lower than that of phase space, the
contraction to which the FT {\it might apply} under the CH is not the full
divergence of the equations of motion: one should rather consider the
contraction of the surface of the attracting set.

3) the fluctuation theorem does not apply to irreversible evolutions, like
$NS_{irr}$, not even if CH holds. 

Results on the determination of the local exponents spectrum in
a $960$ truncation of the $NS_{rev}$ and $NS_{irr}$ equations at high
Reynolds number $R=2048$ are: \*\*

\eqfig{180}{115}{}{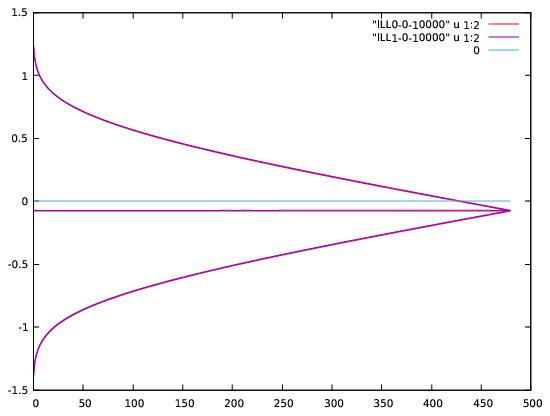}
.\hglue.5cm\raise 2cm \hbox{\kern3.5cm\tiny A}
\vskip-2mm \0{\small Fig.5: The local Lyapunov spectrum in a $960$ modes in
  $NS_{rev}$ and $NS_{irr}$ flows at $R=2048$. The $n=4N(N+1)$ exponents
  $\l_0,\ldots \l_{n-1}$ are drawn reporting for each
  $k=0,\ldots,k_{\frac{n}2-1}$ the values of $\l_k,\l_{n-1-k}$ and the
  average $\frac12(\l_k+\l_{n-1-k})$ for each $k=0,\ldots \frac{n}2-1$
  (``pairing curve'').  The spectra are averaged over a time $800$ units
  sampled every $4$ (quite short): before reaching such times the running
  average values have become stable, although the individual exponents are
  still fluctuating. Also remarkable is the {\it apparent} ``pairing''
  between $\l_k,\l_{n-1-k}$: however this pairing is approximately realized
  only in a range of $R$ and $N$: if $R$ is lowered at fixed $N$ the
  pairing line becomes {\it sensibly curved} (as checked) and the same
  should happen at fixed $R$ and large $N$. Graphs are ``by lines'': but
  also ``by points'' would look continuous because $n$ is large.}  \*
Fig.5 gives the spectra (in the same panel and {\it almost superposed} on the
scale of the drawing) and shows their agreement in corresponding
evolutions. The straight line at level $0$ is a visual aid (it shows
immediately that the sum of the exponents is
$<0$ and that time reversal $I$ {\it is not a symmetry} on the attracting
surface if CH, which implies that motion should be a Anosov flow,x holds).  

Fig.5 exhibits a {\it large number} of observables (\ie the individual
Lyapunov exponents) which, although non local as observables, have the
``same average'' values in corresponding stationary states: namely the
$960$ local Lyapunov exponents in the 2D case of Fig.5 and the PDF's Fig.3
and Fig.4.

Returning to the FR and to the above objections, the latter results on the
Lyapunov spectrum suggest a {\it new viewpoint}.

In \cite{Ga019a,Ga019d} is has been proposed that the first two objections
do not apply to the cases considered here if the following interpretation
of Fig.5 is accepted: the exponents which are part of the negative pairs
have to be discarded being interpreted as the exponents controlling the
uninsteresting attraction by the attracting surface. Hence one remains with
an equal number of positive and negative exponents (\ie only the pairs of
opposite sign count to evaluate the phase space contraction on the
attractor).

The lack of time reversal symmetry applies to the $NS_{rev}$ whenever the
attracting set is smaller than the full phase space (as in the case
reported in Fig.5) and, of course, {\it always} to the $NS_{irr}$. A
different time reversal symmetry mapping the attracting surface into
itself, could be recovered if the assumption that the flow satisfies Axiom
C is accepted, \cite{BG997,Ga013b}.

This has not yet been tested: however the approximate (see caption to
Fig.5) pairing would be very helpful because it suggests
$\sim$proportionality between the sum of the $2n^*$ exponents appearing in
pairs of opposite sign and the sum of all $d=4N(N+1)\defi 2n$ pairs: the
latter is directly accessible from the total divergence and the sum of the
opposite pairs is identified with the phase space contraction of the
attracting set so that average of the latter will simply be
\be\s_{attractor,+}=\frac{num.\ of\
  opposite\ sign\ pairs}{num.\ pairs}\s_+\,\defi\,
\frac{n^*}{n} \s_+\Eq{e7.1}\ee
The contraction $\s_{attractor}$ on the surface of the attracting set at
the configuration $u$ is proposed to be identified with the sum
$\sum_{k=0}^{n^*} (\l_k(u)+\l_{n-k-1}(u))$ of the local exponents.  For a
physical interpretation and relevance in terms of {\it entropy generation} of
$\s_{attractor}$ see \cite[Sec.9]{Ga019d}.

The above comments on problems 1,2,3 could then be tested, at the same
time, by checking validity of FR with slope $\frac{n^*}{n} \tau\s_+$ rather
than $\tau\s_+$: this is a difficult (\ie computationally demanding time),
not an impossible simulation task, but it has not been tested yet.

Further properties of the local Lyapunov spectra and a large scale
representation of the apparent difference between corresponding reversible
and irreversible exponents, is illustrated in the drawings in Fig.6\&7\&8
in the Appendix and in Fig.1\&4, Fig.6\&7, Fig.1\&2 in,
respectively, \cite{Ga019a,Ga019d,Ga019c}.

A question that needs to be studied is whether the Equivalence Hypothesis
extends to 3D, as I conjecture, or more modestly is restricted to 2D; or
whether it demands a $R$ growing with the cut-off, so that the scales of
the local observables are always below the Kolmogorov's scale. As
formulated here, fixed $R$ and a scale above Kolmogorov's scale,
equivalence will be eventually realized if $N$ is large enough (an often
criticized proposal): this is a key difference between the conjecture
formulated in \cite{Ga997b} (for cases with $N$ fixed and $R\to\infty$,
based on strong chaos) and the one here (for cases with $R$ fixed and
$N\to\infty$, based on microscopic chaos and ensembles equivalence).

\kern-5mm
\APPENDICE{1}
\def\SEC{Extra plots}
\section{\SEC}
\label{secA}

Complementary plots illustrate other aspects obtained in
auxiliary simulations.

Remarkably individual local exponents have fluctuations in reversible flows
much larger than those of corresponfing irreversible flows. This is clearly
exhibited in the following figure \vglue3.mm

\eqfig{180}{110}{}{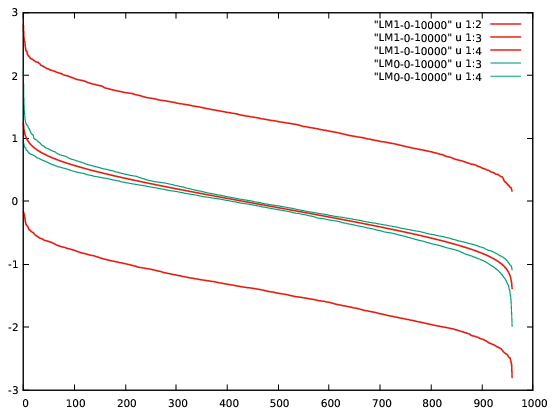}{}
%
\vskip-2mm \0{\small Fig.6: The upper (red o.l.) curve are the loci of the
  largest values observed, in the time $t\le10000$ considered in Fig.5,
  ($960$ modes, $R=2048$), of the {\it reversible flow} exponents; 
  lower (red o.l.) curve are loci of smallest values observed and
  central (red o.l.) line is the actual Lyapunov spectrum for the
  reversible flow (in  Fig.5 curve was drawn
  breaking it in two halves to exhibit pairing: and in the following fig.7
  it is reproduced without breaking it). The two (green o.l.)
  central lines are the upper and lower values observed in the {\it
    irreversible flow} exponents: the drawing shows that the average of the
  reversible flow is between (actually covered by) upper and lower values
  of irreversible flow exponents (whose average values are not drawn
  but on drawing scale would coincide with the reversible flow
  exponents).}

Fig.6 is surprising: the instantaneous local exponents fluctuate very
differently: the reversible ones far more than the irreversible ones but
they have the same averages. And one could think that, drawing all the
instantaneous exponents, in both cases one would fill randomly the space
between the upper fluctuation and the lower one.  Instead the space would
be filled by lines ``parallel'' to the averages.

\eqfig{200}{120}{}{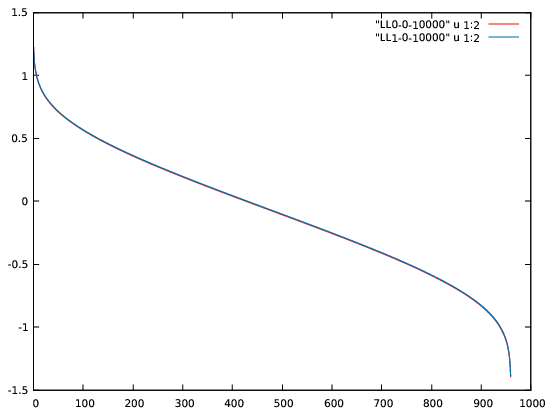}{}

\0{\small Fig.7: Local Lyapunov spectrum in a $960$ modes at $R=2048$ {\it
    for both $NS_{rev}$ and $NS_{irr}$ flows in the same panel},
  overlapping (see however Fig.8). The $n=4N(N+1)$ exponents $\l_0,\ldots
  \l_{n-1}$ for $NS_{rev}$ and $NS_{irr}$ flows are drawn and are
  apparently {\it superposed}.  Spectra are averaged over a time $4*3600$
  units of $4/h$ steps of size $h=2^{-17}$, sampled every $4$: before
  reaching such times the running average values have become stable,
  although the individual exponents are still fluctuating. See Fig.6
  above.}

However the two spectra, overlapping in Fig.7, differ
as shown in the Fig.8 below.

\eqfig{200}{120}{}{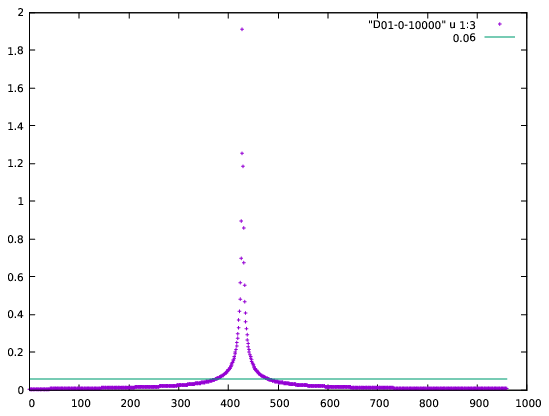}{}

\0{\small Fig.8: The two spectra in the previous figure are here
individually compared term by term, drawing for each $k\in [0,960)$
the difference $\frac{|\l_k^{irr}-\l_k^{rev}|}{(|\l_k^{irr}|+|\l^{rev}|)/2}$.
The line marks $6\%$. The larger relative difference at the center of the
spectrum mostly reflects that it is there that the exponents are close to
zero so that the numerical errors are larger.}

Preliminary results on the local Lyapunov spectrum in a $3968$ modes
truncation are in the graph in Fig.8: this is a difficult computation due to the
computer time necessary.

\eqfig{180}{120}{}{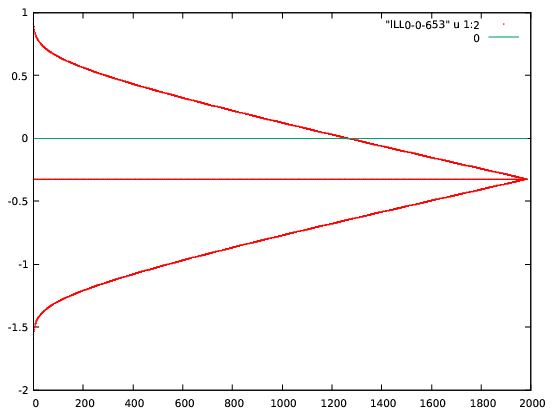}{}

\0{\small Fig.9: The $NS_{irr}$ local Lyapunov spectrum as in Fig.2 but for
  a large truncation ($3968$ modes). Still shows a rather clear
  (approximate) pairing. The same spectrum for the $NS_{rev}$ case is close
  although appreciably different on the drawing scale: the problem is that
  the number of modes is very large and the exponents are averaged over a
  relatively small time span ($t=800$ while in Fig.7 it is $t=10000$) due
  to the computer time need.}

\APPENDICE{2}
\def\SEC{\ Irreversible and reversible 3D NS}
\section{\SEC}
\label{secB}

In dimension $d=3$ the velocity field is represented as

\be\eqalign{
\uu(\xx)=&\sum_{\V0\ne\kk\in Z^3} \uu_\kk
  e^{-i\kk\cdot\xx},\quad
\uu_\kk=\sum_{\th=\pm1} u_{\th,\kk} e_\th(\kk)\cr}
\Eq{aB.1}\ee
with $u_{\kk,\th}=\lis u_{-\kk,\th}$ scalars,
and $e_{\th}(\kk)=e_{\th}(-\kk),\th=\pm1$ are two mutually orthogonal unit
vectors in the plane orthogonal to $\kk$: $e_\th(\kk)\cdot                                              
 e_{\th'}(\kk)=\d_{\th,\th'}$.

The Euler equation for the components $u_{\th,\kk}$ is then
\be\eqalign{&\dot{\uu}_\kk=\EE(\uu)_\kk, \qquad \EE(\uu)_{\th,\kk}=\cr
&\kern-3mm\tiny\sum_{\kk_1+\kk_2=\kk}
(i\kk_2\cdot e_{\th_1}(\kk_1))(e_{\th_2}(\kk_2)\cdot e_\th(\kk))
u_{\th_1,\kk_1}u_{\th_2,\kk_2}
\cr}\Eq{aB.2}\ee
The helicity conservation follows by checking that $\frac{d}{dt}                                        
\int \uu(\xx)\cdot(\BDpr\wedge \uu(\xx))\,d\xx=0$.

The reversible version of the $3$-dimensional NS equations, in which
enstrophy is a constant of motion, is
$\dot\uu_\kk=\EE_\kk(\uu)-\a(\uu)\kk^2\uu_\kk+\V f_\kk$ with
$\a$ defined by:

\be\eqalign{
&\a(\uu)=\frac{\big(\sum\scriptstyle
\kk^2\,(\uu_{\kk_1}\cdot i\kk_2)(\uu_{\kk_2}\cdot
\uu_{-\kk})\big) + \big(\sum_\kk \kk^2 f_\kk\cdot\uu_{-\kk}\big)}
{\scriptstyle\sum_\kk \kk^4 |\uu_\kk|^2}\cr}
\Eq{aB.3}\ee
where the firs sum runs over $\kk_1,\kk_2,\kk$ with $\kk=\kk_1+\kk_2$.
\*

\0{\it Acknowledgements:} {\it I am grateful to L.Biferale (Roma2),
L.S.Young (CIMS), and to L.Silvestrini (Roma1-INFN) for support and
for making available use of their clusters.}

\bibliographystyle{plain} 


\*
\0email: {\tt giovanni.gallavotti@roma1.infn.it}

\0web:   {\tt https://ipparco.roma1.infn.it}

\vfill\eject
.

\vfill\eject
.\raise 6cm \hbox{\centerline{\bf Fig.s 1-9 scaled}}

\vglue1cm

\eqfig{160}{115}{}{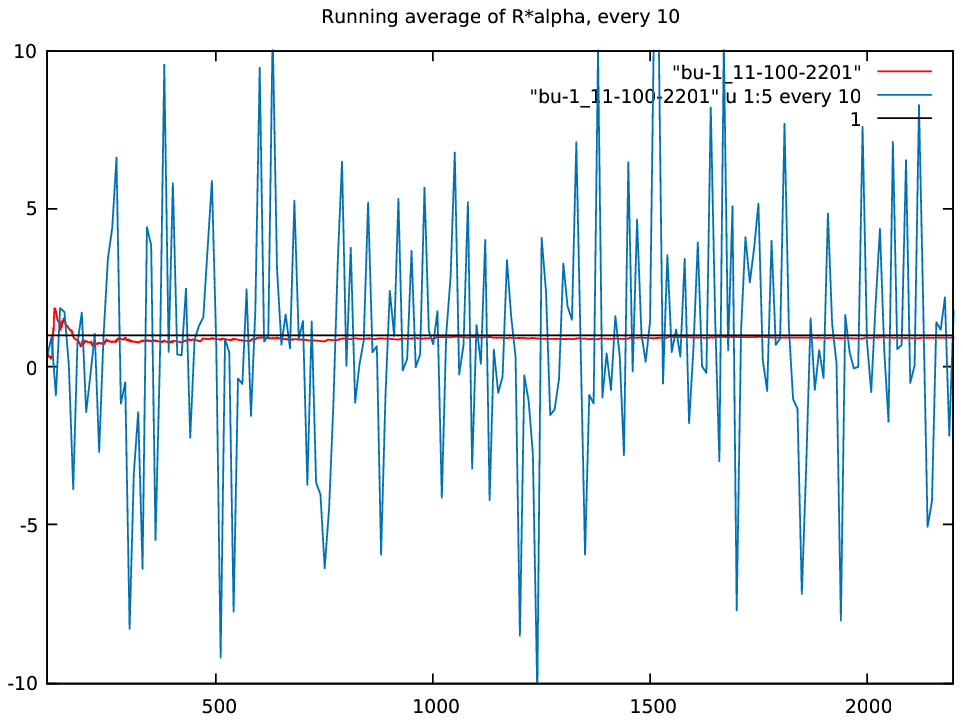}{}


\vglue6cm
\eqfig{160}{115}{}{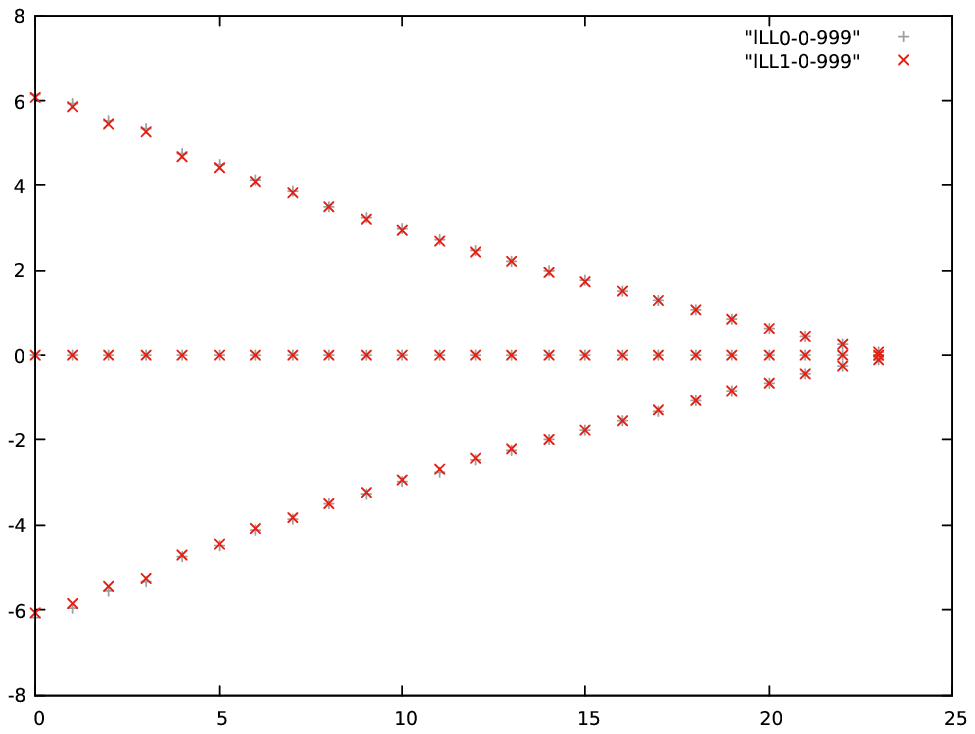}{}
\vfill\eject
.
\vfill\eject
\vglue3cm
\eqfig{160}{130}{}{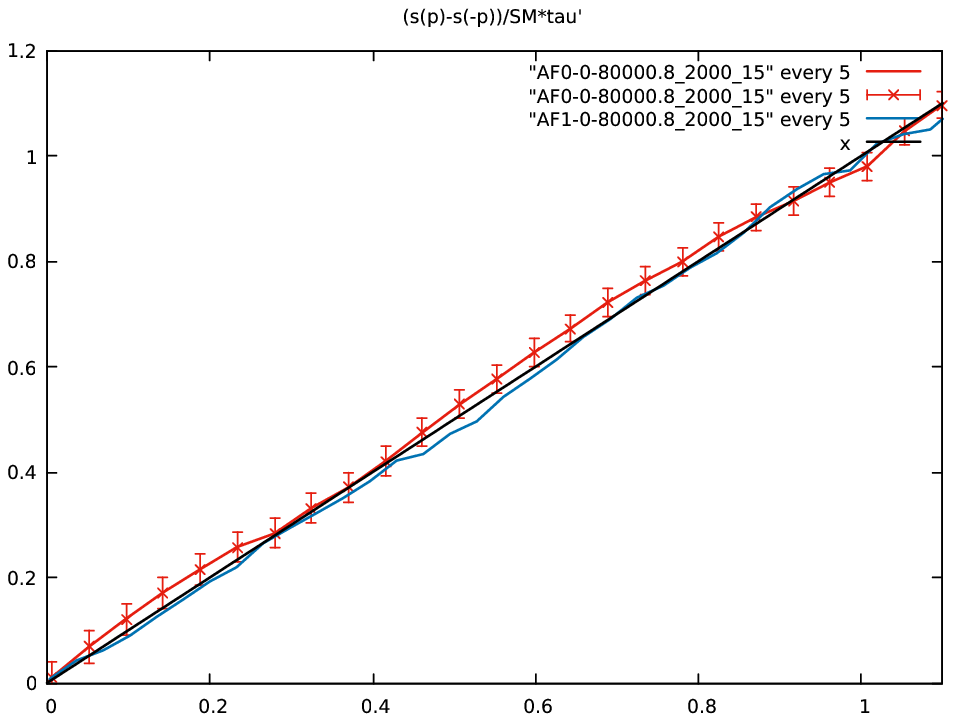}{}
\vglue6cm
\eqfig{160}{115}{}{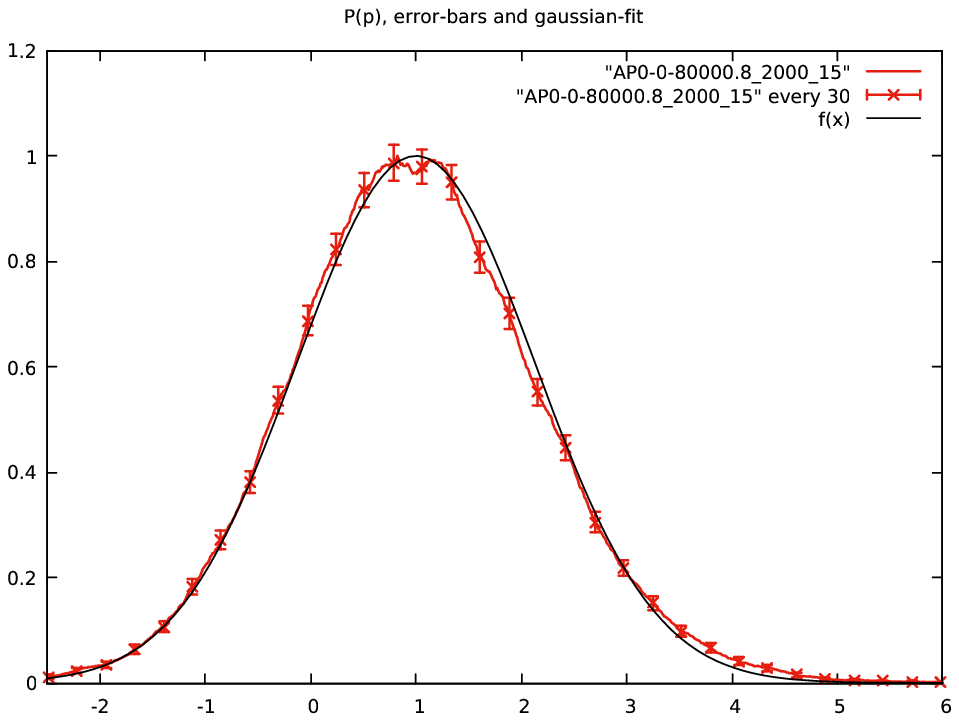}{}
\vfill\eject
.
\vfill\eject
\vglue4cm
\eqfig{160}{110}{}{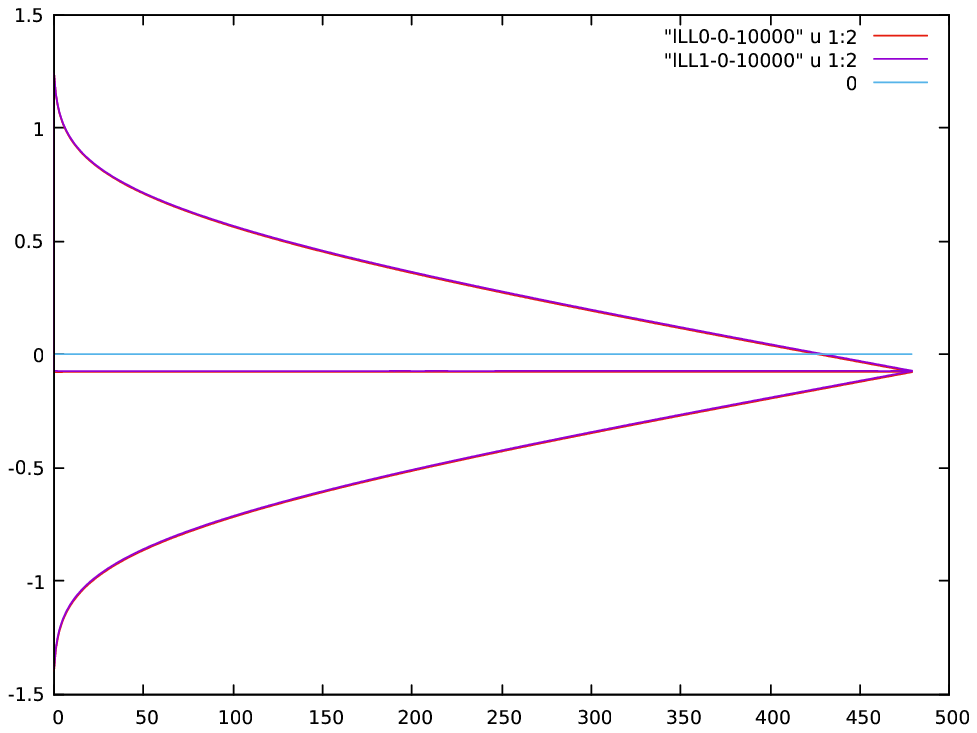}{}

\vglue6cm
\eqfig{160}{110}{}{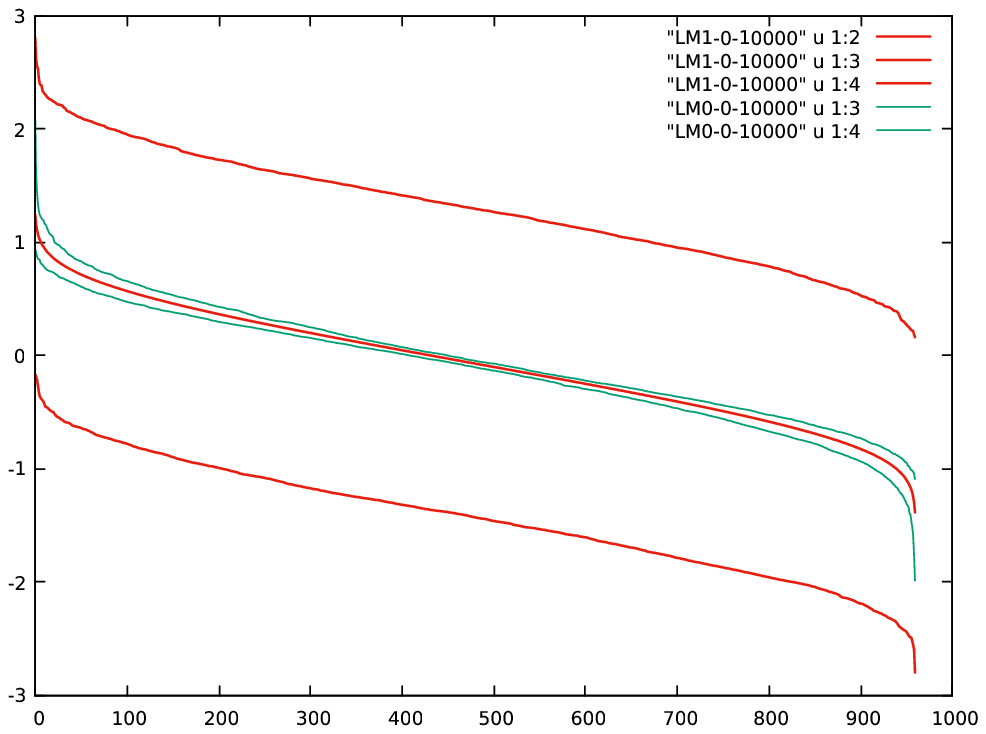}{}
\vfill\eject
.
\vfill\eject

\vglue4cm
\eqfig{160}{110}{}{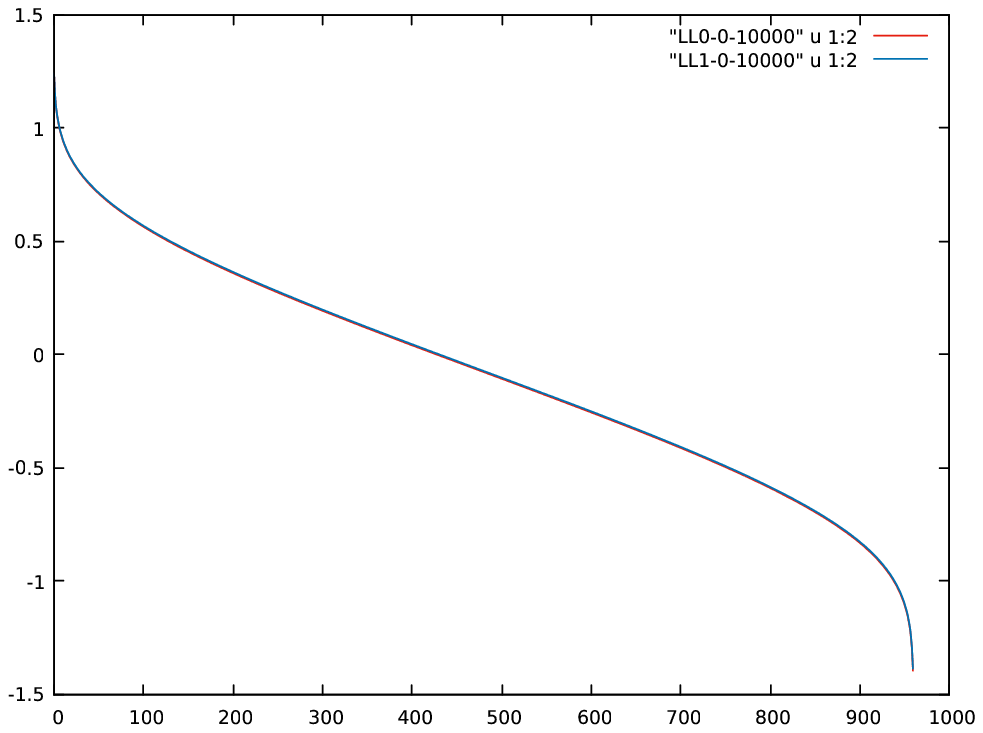}{}

\vglue6cm
\eqfig{160}{110}{}{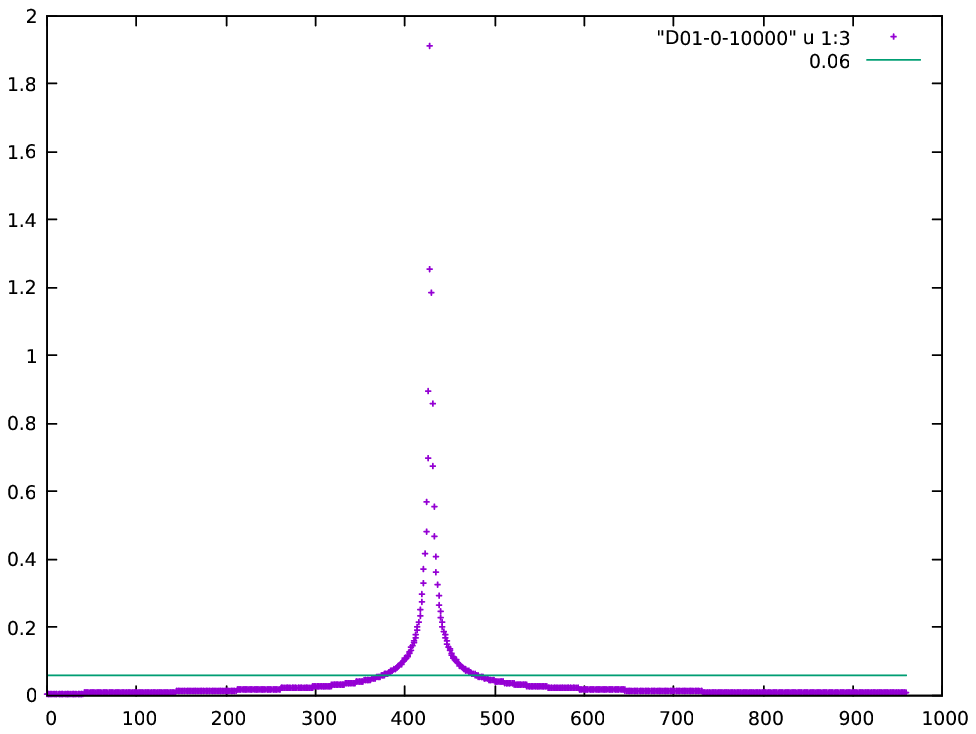}{}
\vfill\eject
.
\vfill\eject
\vglue6cm
\eqfig{660}{110}{}{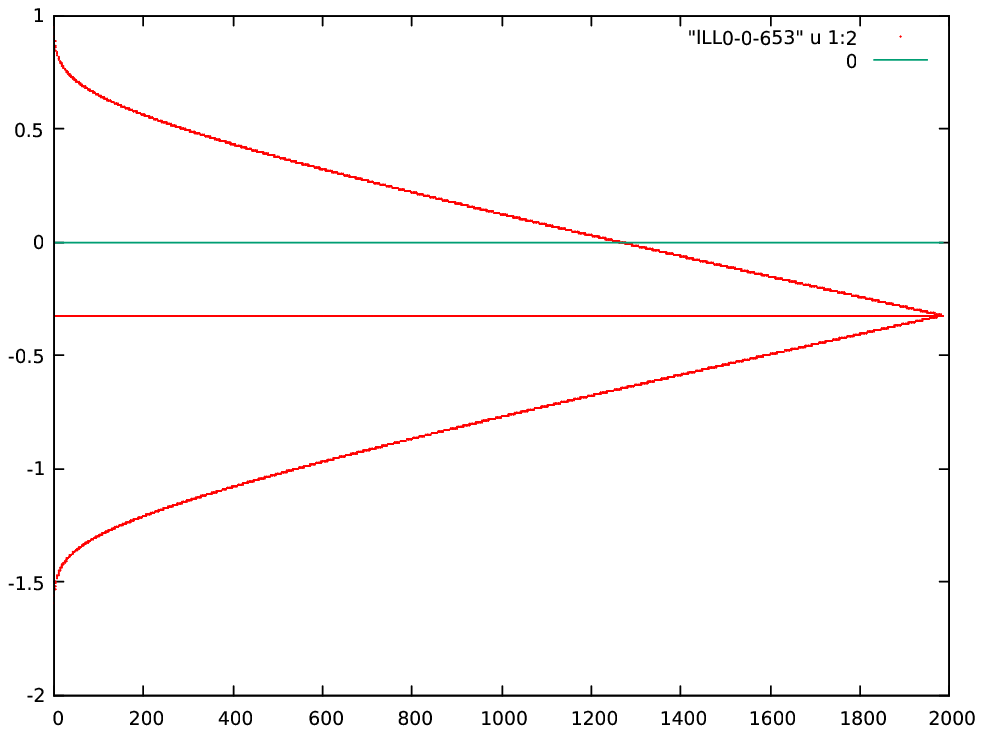}{}
\end{document}